\documentstyle[jkas2]{article} 
\runningauthor{KANG} 
\runningtitle{Cosmic Ray Acceleration at SNRs} 
\beginpage{1} 
\endpage{11} 

\def\etal{{\it et al.~}}
\def\eg{{\it e.g.,~}}
\def\ie{{\it i.e.,~}}
\def\kms{~{\rm km~s^{-1}}}
\def\cm3{~{\rm cm^{-3}}}

\def\lsim{\mathrel{  
        \raise0.3ex\hbox{$<$}\kern-0.75em{\lower0.65ex\hbox{$\sim$}}}}
\def\gsim{\mathrel{
        \raise0.3ex\hbox{$>$}\kern-0.75em{\lower0.65ex\hbox{$\sim$}}}}

\begin{document}

\title{COSMIC RAY ACCELERATION AT BLAST WAVES FROM TYPE Ia SUPERNOVAE}

\author{HYESUNG KANG}

\address{Department of Earth Sciences, Pusan National University, Pusan  609
-735, Korea \\
{\it E-mail: hskang@pusan.ac.kr }}

\address{\normalsize{\it (Received November 10, 2006; Accepted December 6, 2006)}}

\abstract{
We have calculated the cosmic ray (CR) acceleration at young
remnants from Type Ia supernovae expanding into a uniform interstellar medium
(ISM).
Adopting quasi-parallel magnetic fields, gasdynamic equations 
and the diffusion convection equation for the particle distribution function
are solved in a comoving spherical grid which expands with the shock.
Bohm-type diffusion due to self-excited Alfv\'en waves, 
drift and dissipation of these waves in the precursor and   
thermal leakage injection were included.  
With magnetic fields amplified by the CR streaming instability, 
the particle energy can reach up to $10^{16}Z$ eV at young
supernova remnants (SNRs) of several thousand years old. 
The fraction of the explosion
energy transferred to the CR component asymptotes to 40-50 \% by that time. 
For a typical SNR in a warm ISM, the accelerated CR energy spectrum should 
exhibit a concave curvature with the power-law slope flattening 
from 2 to 1.6 at $E\gsim 0.1$ TeV. 
}

\maketitle

\section{INTRODUCTION}

Diffusive shock acceleration (DSA) is widely
accepted as the primary mechanism through which cosmic rays are
generated in a wide range of astrophysical shocks (Drury, 1983; Malkov \&
Drury 2001 and references therein; Kang \& Jones 2002; Kang 2003).
It is well known that the CR energy density is comparable to the gas
thermal energy density in the ISM and plays important
dynamical roles in the evolution of our Galaxy.
Most of galactic cosmic rays, at least up to $10^{14}$ eV of the
proton energy, are believed to be accelerated by SNRs 
within our Galaxy via Fermi first order process 
(Blanford \& Ostriker 1978; Lagage \& Cesarsky 1983; Blandford \& Eichler 1987;
Drury \etal 2001). 

Time-dependent, kinetic simulations of the CR acceleration at SNRs
have shown that up to 50 \% of explosion energy can be converted to CRs, 
when a fraction $10^{-4}-10^{-3}$ of incoming thermal particles are injected
into the CR population at the subshock 
(\eg Berezhko, Ksenofontov, \& Yelshin 1995;
Berezhko, \& V\"olk 1997;  V\"olk \& Berezhko 2005; Kang \& Jones 2006).
This should be enough to replenish the galactic CRs escaping from our Galaxy
with $L_{CR}\sim 10^{41} {\rm erg ~s^{-1}}$.
X-ray observations of young SNRs such as SN1006 and Cas A
indicate the presence of TeV electrons emitting nonthermal
synchrotron emission immediately inside the outer SNR shock
(Koyama \etal 1995; Bamba \etal 2003).
They provide clear evidences for the efficient acceleration of the CR electrons 
at SNR shocks.   
Also recent HESS observation of SNR RXJ1713.7-3946
indicates possible detections of $\pi^0$ decay $\gamma$ rays
from the hadronic CRs accelerated by the SNR shock
(Aharonian \etal 2004; Berezhko \& V\"olk 2006). 

In the kinetic equation approach to numerical study of CR acceleration 
at shocks, the diffusion-convection equation for the particle momentum 
distribution, $f(p)$, is solved along with suitably modified gasdynamic 
equations (\eg Kang \& Jones 1991).
Accurate solutions to the CR diffusion-convection equation
require a computational grid spacing significantly smaller 
than the particle diffusion length, $\Delta x \ll x_d(p) = \kappa(p)/u_s$,
where $\kappa(p)$ is diffusion coefficient and $u_s$ is the shock speed.
In a realistic diffusion transport model,
the diffusion coefficient has a steep momentum dependence, 
$\kappa (p) \propto p^s$, with $s \sim 1-2$.
So a wide range of length scales is required to be resolved 
in order to follow the CR acceleration from the injection energy 
(typically $p_{\rm inj}/m_{\rm p}c \sim 10^{-2}$) to highly relativistic 
energy ($p/m_{\rm p}c \gg 1$). This constitutes an extremely challenging
numerical task, requiring rather extensive computational resources.

To overcome this numerical problem, Berezhko and collaborators
(\eg Berezhko \etal 1995) introduced a
``change of variables technique'' in which the radial coordinate is
normalized by the diffusion length, $x_d(p)$, at each particle 
momentum for the upstream region. 
This allowed them to solve the coupled system of gasdynamic
equations and the CR transport equation with $\kappa(p)\propto p$.
Their scheme was designed for simulations of supernova remnants, 
which were represented by piston-driven spherical shocks 
in one-dimensional (1D) spherical geometry.

On the other hand,
Kang and collaborators have taken an alternative approach that
is based on a more conventional Eulerian formalism.
Adaptive Mesh Refinement (AMR) technique and subgrid shock tracking 
technique were combined to build CRASH (Cosmic-Ray Amr SHock) code 
in 1D plane-parallel geometry (Kang \etal 2001) and in 1D spherical
symmetric geometry (Kang \& Jones 2006).
In order to implement the shock tracking and AMR techniques effectively
in a spherical geometry, we solve the fluid and diffusion-convection 
equations in a frame comoving with the outer spherical shock.
Adopting a comoving frame turns out to be a great numerical success,
since we can achieve numerical convergence at a grid resolution
much coarser than that required in an Eulerian grid.
In the comoving grid, the shock remains at the same location,
so the compression rate is applied consistently to the CR distribution
at the subshock, resulting in much more accurate and efficient 
low energy CR acceleration.

In the present paper, we apply the spherical CRASH code for the
CR acceleration at remnant shocks from Type Ia SNe expanding into a uniform
interstellar medium, assuming a quasi-parallel field geometry.
Details of the numerical method are described in \S II.
The simulation results are presented and discussed in \S III,
followed by a summary in \S IV.

\section{NUMERICAL METHOD}

\begin{figure*}[t]
\vskip -0.5cm
\centerline{\epsfysize=14cm\epsfbox{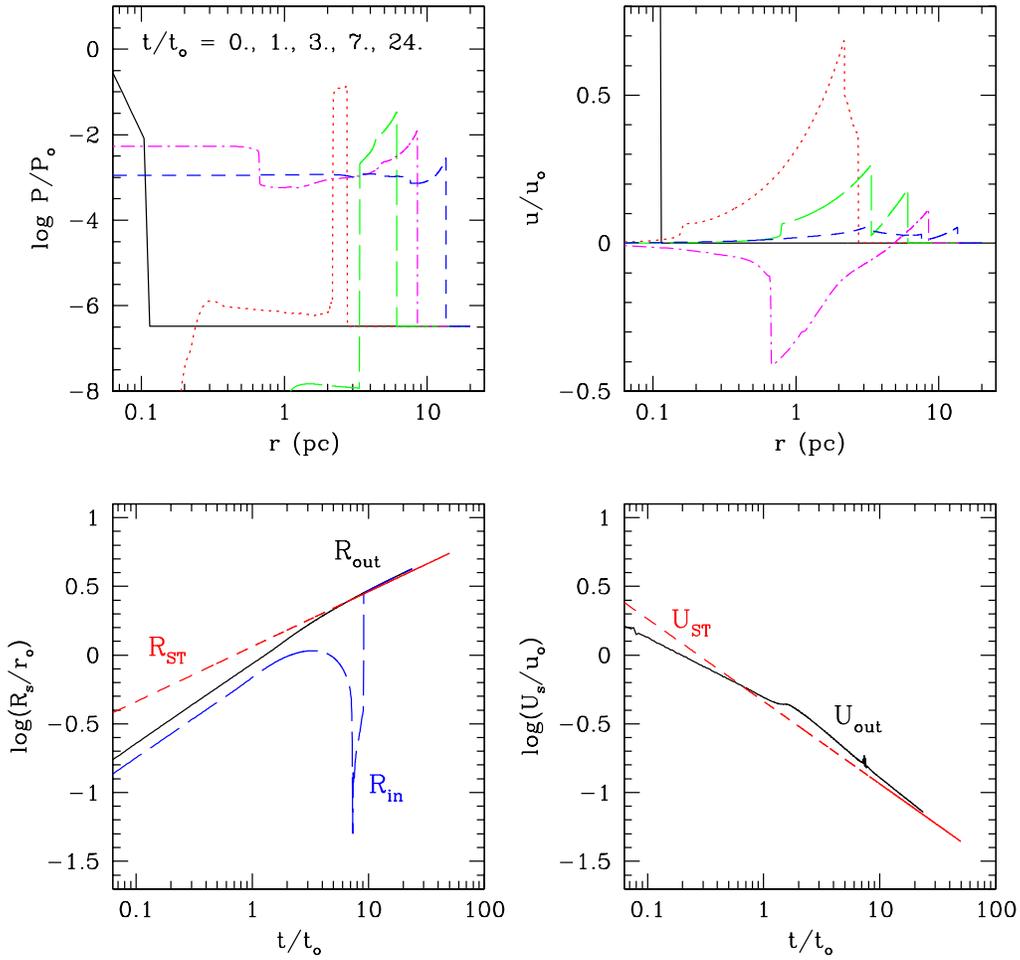}}
\vskip -0.5cm
\caption{
Top two panels show the evolution of a typical SNR in a gasdynamic
simulation.
Bottom two panels show the positions of the outer forward shock
($R_{\rm out}$) and inner reverse shock ($R_{\rm in}$) and the speed of
the outer shock ($U_{\rm out}$) as a function of time.
The Sedov Taylor similarity  solutions for the outer shock, $R_{\rm ST}$ and 
$U_{\rm ST}$ are also shown for comparison.
}
\end{figure*}

\subsection{BASIC EQUATION}

Here we consider the CR acceleration at a quasi-parallel shock 
where the magnetic field lines are parallel to the shock normal. 
So we solve the standard gasdynamic equations with CR pressure terms
added in the Eulerian formulation for one dimensional spherical 
symmetric geometry.

\begin{equation}
{\partial \rho \over \partial t}  +  {\partial\over \partial r} (\rho u)
= -{ 2 \over r} \rho u,
\label{masscon}
\end{equation}

\begin{equation}
{\partial (\rho u) \over \partial t}  +  {\partial\over \partial r} (\rho u^2 +
P_g + P_c)
= -{2 \over r} \rho u^2,
\label{mocon}
\end{equation}

\begin{equation}
{\partial (\rho e_g) \over \partial t} + {\partial \over \partial r}
(\rho e_g u + P_g u) =
-u {{\partial P_c}\over {\partial r}}
-{2 \over r} (\rho e_g u + P_g u),
\label{econ}
\end{equation}

\begin{equation}
{\partial S\over \partial t}  +  {\partial\over \partial r} (S u) =
-{2\over r} S u
+ {(\gamma_{\rm g} -1)\over \rho^{\gamma_{\rm g} -1} } \left[W(r,t) - L(r,t)\right],
\label{scon}
\end{equation}

where $P_{\rm g}$ and $P_{\rm c}$ are the gas and the CR pressure,
respectively, $e_{\rm g} = {P_{\rm g}}/{[\rho(\gamma_{\rm g}-1)]}+ u^2/2$
is the total energy of the gas per unit mass.
The evolution of a modified entropy, $S = P_g/\rho^{\gamma_g - 1}$,
is followed everywhere except across the subshock,
since for strongly shocked flows
numerical errors in computing the gas pressure from the total
energy can lead to spurious entropy generation with
standard methods, especially in the shock precursor
(Kang, Jones, \& Gieseler 2002). 
Total energy conservation is applied only across the subshock.
The remaining variables, except for $L$ and $W$, have standard meanings.
The injection energy loss term, $L(r,t)$, accounts for the
energy carried by the suprathermal particles injected into the CR component at
the subshock.
Gas heating due to Alfv\'en wave dissipation in the upstream region is
represented by the term
\begin{equation}
W(r,t)= - v_A {\partial P_c \over \partial r },
\end{equation}
where $v_A= B/\sqrt{4\pi \rho}$ is the Alfv\'en speed.
This term derives from a simple model in which Alfv\'en waves are amplified by
streaming CRs and dissipated locally as heat in the precursor region
(\eg Jones 1993).

The CR population is evolved by solving the diffusion-convection equation,
\begin{eqnarray}
{\partial g\over \partial t}  + (u+u_w) {\partial g \over \partial r}
= {1\over{3r^2}} {\partial \over \partial r} \left[r^2 (u+u_w)\right] 
\left( {\partial g\over \partial y} -4g \right) \nonumber\\ 
+ {1 \over r^2}{\partial \over \partial r} \left[r^2 \kappa(r,y)
{\partial g \over \partial r}\right],
\label{diffcon}
\end{eqnarray}
where $g=p^4f$, with $f(p,r,t)$ the pitch angle averaged CR
distribution,
and $y=\ln(p)$, while $\kappa(r,y)$ is the diffusion coefficient
parallel to the field lines
(Skilling 1975).
For simplicity we express the particle momentum, $p$ in
units $m_{\rm p}c$ hereafter
and consider only the proton CR component.
The wave speed is set to be $u_w=v_A$ in the upstream region, while we
use $u_w=0$ in the downstream region.
This term reflects the fact that
the scattering by Alfv\'en waves tends to isotropize
the CR distribution in the wave frame rather than the gas frame.

\begin{table*}
\begin{center}
{\bf Table 1.}~~Model Parameters\\
\vskip 0.3cm
\begin{tabular}{ lrrrrrrrr }
\hline\hline
Model & $n_{\rm ISM}$ & $E_o$ & ~$B_{\mu}$ & $\epsilon_B$ & $r_o$ &~~$t_o$ & 
  $u_o$ &$P_o$  \\
~ & $(cm^{-3})$ & ($10^{51}$ ergs) & $\mu G$ & & (pc) & (years) & 
($10^4 \kms$) & ($10^{-6}$erg cm$^{-3}$) \\

\hline
S1  & 0.3   &1 & 30 & 0.16& 3.19 & 255. &1.22 & 1.05 \\
S2  & 0.3   &4 & 30 & 0.16& 3.19 & 127. &2.45 & 4.20 \\
S3  & 0.3   &1 &  5 & 0.16& 3.19 & 255  &1.22 & 1.05 \\
S4  & 0.003 &1 & 30 & 0.16& 14.8 & 1182. &1.22 & 1.05~(-2)\\
S5  & 0.3   &1 &  5 & 0.2& 3.19 & 255  &1.22& 1.05 \\

\hline
\end{tabular}
\end{center}
\end{table*}

\subsection{Spherical CRASH code}
Details of the CRASH code in 1D spherical symmetric geometry
can be found in Kang \& Jones (2006),
so we briefly describe the basic features here. 
We solve the equations (1)-(6) in a comoving frame that expands 
with the instantaneous shock speed, 
since a spherical shock can be made to be stationary 
by adopting comoving variables which factor out a uniform 
expansion or contraction.
Because the shock is at rest and tracked accurately
as a true discontinuity, we can refine the region around
the gas subshock at an arbitrarily fine level.
The AMR technique allows us to ``zoom in'' inside the precursor structure
with a hierarchy of small, refined grid levels applied around the shock.
The result is an enormous savings in both computational time and
data storage over what would be required to solve the problem using
more traditional methods on a single fine grid.

\begin{figure*}[t]
\vskip -0.5cm
\centerline{\epsfysize=14cm\epsfbox{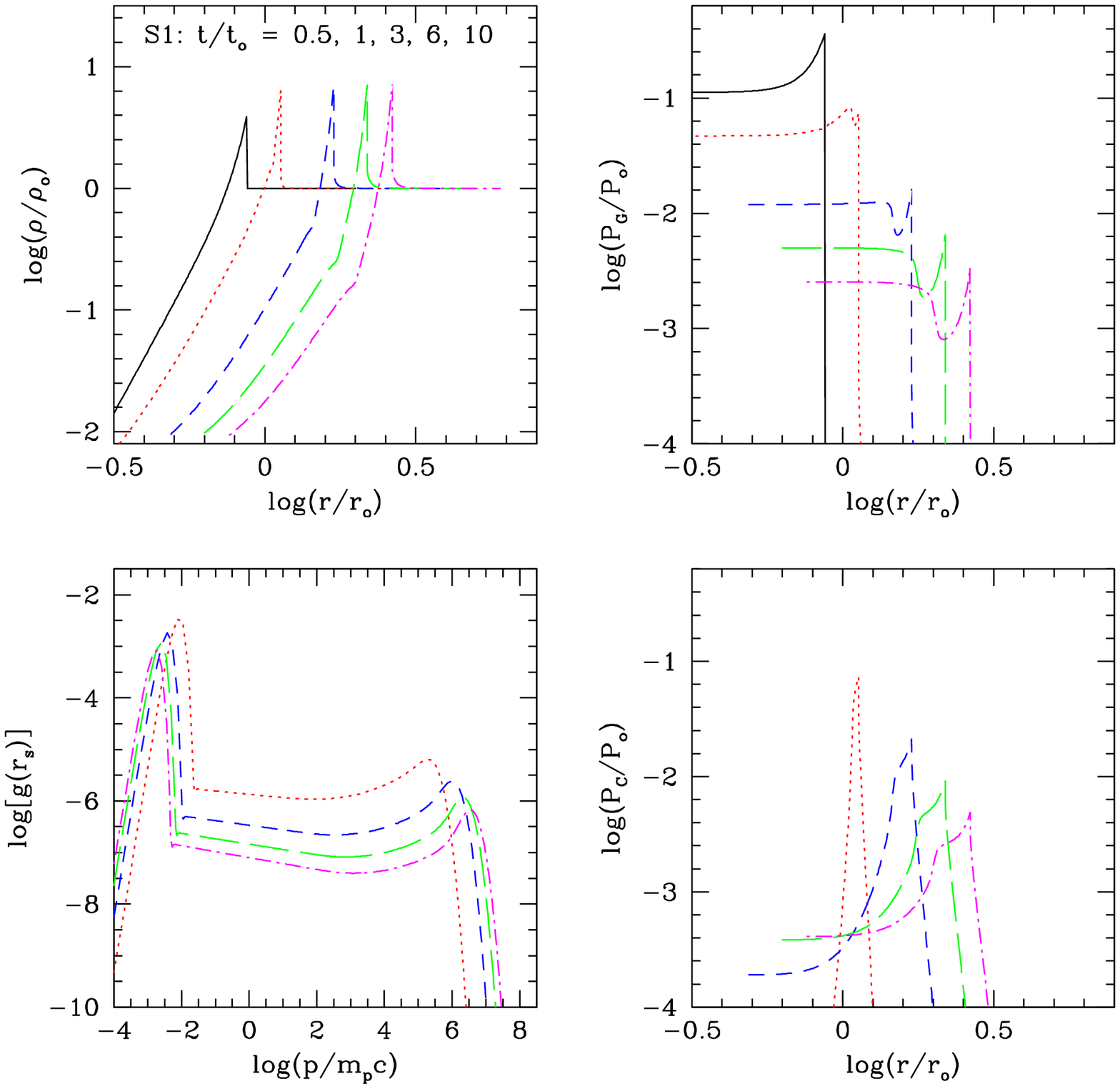}}
\vskip -0.5cm
\caption{
Evolution of model S1 SNR expanding into a uniform ISM
at $t/t_o = 0.5, ~1.,~6.,$ and 10.
The model parameters are
$M_{ej}=1.4 M_{\sun}$, $E_o=10^{51}$ ergs, $n_{\rm ISM}=0.3 {\rm cm}^{-3}$,
and $B_{\mu}=30$.
The injection parameter for thermal leakage injection, $\epsilon_B=0.2$.
The lower left panel shows the volume integrated CR spectrum,
$G(p) = \int f(r,p)p^4 r^2 {\rm dr} $.
The initial condition at $ t/t_o=0.5$ (solid line) is set by the
Sedov-Taylor similarity solution.
}
\end{figure*}

\subsection{The Thermal Leakage Injection Model}
The injection rate with which suprathermal particles are injected into CRs
at the subshock depends in general upon the shock Mach number, 
field obliquity angle, and strength of Alfv\'en turbulence responsible 
for scattering.
The CRASH codes treat this process naturally and self-consistently via 
``thermal leakage'' through lowest momentum bins.
The thermal leakage injection model emulates the process by which
suprathermal particles well into the tail of the postshock Maxwellian
distribution leak upstream across a quasi-parallel shock 
(Malkov \& V\"olk 1998; Malkov \& Drury 2001).
This filtering process is implemented numerically by
adopting a ``transparency function'', $\tau_{\rm esc}(\epsilon_B, \upsilon)$,
that expresses the probability of supra-thermal
particles at a given velocity, $\upsilon$, leaking upstream through the
postshock MHD waves (Kang \etal 2002).
One free parameter controls this function; namely, $\epsilon_B = B_0/B_{\perp}$,
which is the inverse ratio of the amplitude
of the postshock MHD wave turbulence $B_{\perp}$ to the general magnetic field
aligned with the shock normal, $B_0$ (Malkov \& V\"olk 1998).
Plasma hybrid simulations and theory both suggest that
$0.25 \lesssim \epsilon_B \lesssim 0.35$, 
so that the model is well constrained.
However, such large values of $\epsilon_B$ lead to very
efficient initial injection and 
most of the shock energy is quickly transferred to 
the CR component for strong shocks considered here ($50<M_s<300$ initially),
causing a numerical problem at the very early stage of simulations.
So we adopted smaller values, $\epsilon_B=0.16-0.2$ in this study.
Dependence of the CR injection and acceleration on this parameter will be
discussed below.

\subsection{A Bohm-like Diffusion Model}

Self-excitation of Alfv\'en waves by the CR streaming instability 
in the upstream region is an integral part of the DAS at SNRs 
(Bell 1987; Vo\"lk \etal 1988; Lucek \& Bell 2001). 
The particles are resonantly scattered by those waves, diffuse
across the shock, and get injected into the Fermi first-order process.
These complex interactions are represented by the diffusion 
coefficient, which is expressed
in terms of a mean scattering length, $\lambda$, and the particle
speed, $\upsilon$, as
$\kappa(x,p) =  \lambda \upsilon/3$.
The Bohm diffusion model is commonly used to represent a saturated
wave spectrum (\ie $\lambda = r_g$, where $r_g$ is the gyro-radius), 
$\kappa_B(p) =  \kappa_n p^2/\ (p^2+1)^{1/2}$.
Here 
$\kappa_n= m c^3/(3eB)= 3.13\times 10^{22} {\rm cm^2s^{-1}} B_{\mu}^{-1}$,
and $B_{\mu}$ is the magnetic field strength in units of microgauss.
Because of the steep momentum dependence for non-relativistic particles
($p\ll 1$), simulations with a Bohm diffusion model
require extremely fine grid resolution around the shock where freshly
injected CRs are concentrated.
Instead we adopt a Bohm-like diffusion coefficient
that includes a weaker non-relativistic momentum dependence,
\begin{equation}
\kappa(r,p) = \kappa_{\rm n}\cdot p \left({\rho_0 \over \rho(r)}\right).
\end{equation}
Previous studies showed that simulations using these two types of
diffusion coefficient produced very similar results
(Kang \etal 2001).
The assumed density dependence for $\kappa$ accounts for compression of the
perpendicular component of the wave magnetic field and
also inhibits the acoustic instability in the precursor of highly modified
CR shocks (Kang, Jones, \& Ryu, 1992).
Hereafter we use the subscripts '0', '1', and '2' to denote
conditions far upstream of the shock, immediately upstream of the
gas subshock and immediately downstream of the subshock, respectively.

\section{Simulations of Sedov-Taylor Blast Waves}

\begin{figure*}[t]
\vskip -0.5cm
\centerline{\epsfysize=14cm\epsfbox{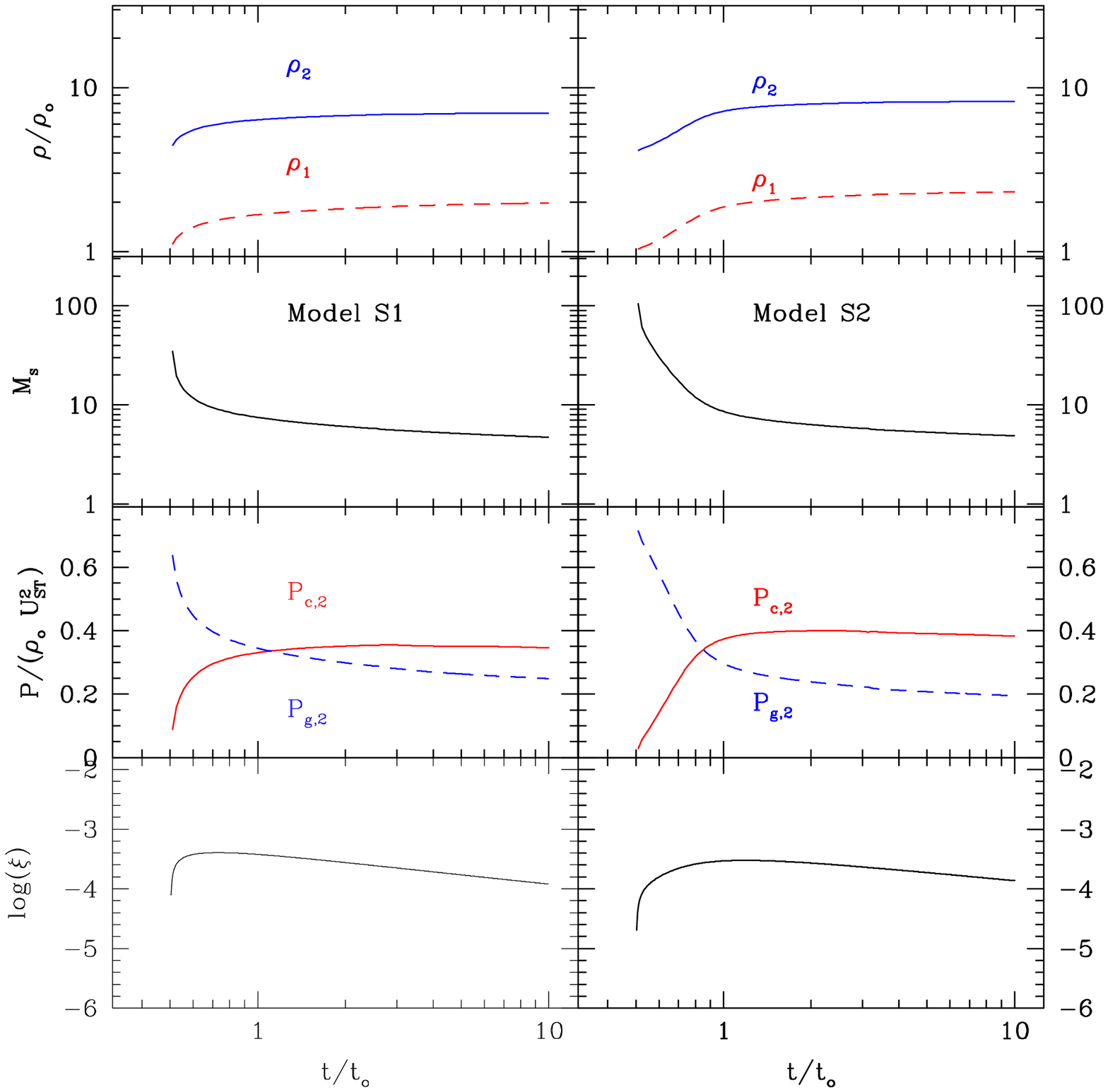}}
\vskip -0.5cm
\caption{
Immediate pre-subshock density, $\rho_1$, post-subshock density,
$\rho_2$, shock Mach number, $M_s$, post-subshock CR and gas pressure
in units of the ram pressure of the unmodified Sedov-Taylor solution,
$\rho_0 U_{ST}^2 \propto (t/t_o)^{-6/5}$, and the CR injection parameter, 
$\xi$, are plotted for models S1 and S2.
}
\end{figure*}

For a supernova remnant propagating into a uniform ISM, 
the CR acceleration takes place mostly during free expansion
and Sedov-Taylor (ST hereafter) stages,
since the shock slows down significantly afterward.
Fig. 1 shows the evolution of a typical SNR calculated by
a hydrodynamics code without the CR pressure terms.
This demonstrates that the ST solution is established
only after the inner reverse shock is reflected at the center 
at $t/t_o \sim 7$ 
(see below for the definition of normalization constants).
Before that time, the reverse shock is
strong and dynamically important.
In our simulations, however, we will ignore the reverse shock,
because the current version of CRASH code can treat only
one shock.
Application of our AMR algorithm for multiple spherical shocks is
not simple, since it requires multiple, comoving grids.
The CR acceleration at the reverse shock is thought to be not
important, because the kinetic energy passed through the reverse 
shock is relatively small.  
Also adiabatic losses by CRs accelerated early on in the interior and then
advected outward through the ST phase would generally be very large.

On the other hand, Fig. 1 shows that the evolution of 
the outer shock speed, $U_{\rm out}(t)$ can be approximated by
the Sedov solution $U_{\rm ST}\propto (t/t_o)^{-2/5}$ for $t/t_o>0.2$.
In order to take account for the CR acceleration from free expansion
stage through ST stage, we begin the calculations with the 
ST similarity solution at $t/t_o=0.5$.
In principle we could start the simulations from $t/t_o \sim 0.1$. 
Such simulations, however, require very long computational time,
since we need to include the extremely hot region with fast sound speeds
near the explosion center. 
We carried out one model from $t/t_o=0.2$ and compared it 
with the case started from $t/t_o=0.5$ in the next section. 
Since the total CR energy gain is proportional to the kinetic
energy passed through the shock, 
$E_{sw}= 2\pi \int \rho_0 U_{\rm out}^3 R_{\rm out}^2 dt$,
we expect our calculations should capture the key aspects of the
CR acceleration at the outer SNR shock.

We terminated the simulation at $t/t_o=10$, while the ST stage
ends typically when the shock slows down to $U_{\rm ST} < 300 \kms$
around $t/t_o\sim 3000$.  However, it has been shown that
the highest momentum, $p_{\rm max}$, is achieved by the end of the 
free expansion stage and the transfer of explosion
energy to the CR component occurs mostly during the early ST stage
(\eg Berezhko \etal 1997).
We will show in the next section that the shock properties and
the CR acceleration efficiency reach roughly time asymptotic values 
for $t/t_o > 1$.
Also it was suggested that non-linear wave damping and the wave dissipation 
due to ion-neutral collisions may suppress the MHD waves significantly 
at the late ST stage, leading to fast particle diffusion and inefficient
acceleration. 
(V\"olk \etal 1998; Ptuskin \& Zirakashvili 2003)

\begin{figure*}[t]
\vskip -0.5cm
\centerline{\epsfysize=14cm\epsfbox{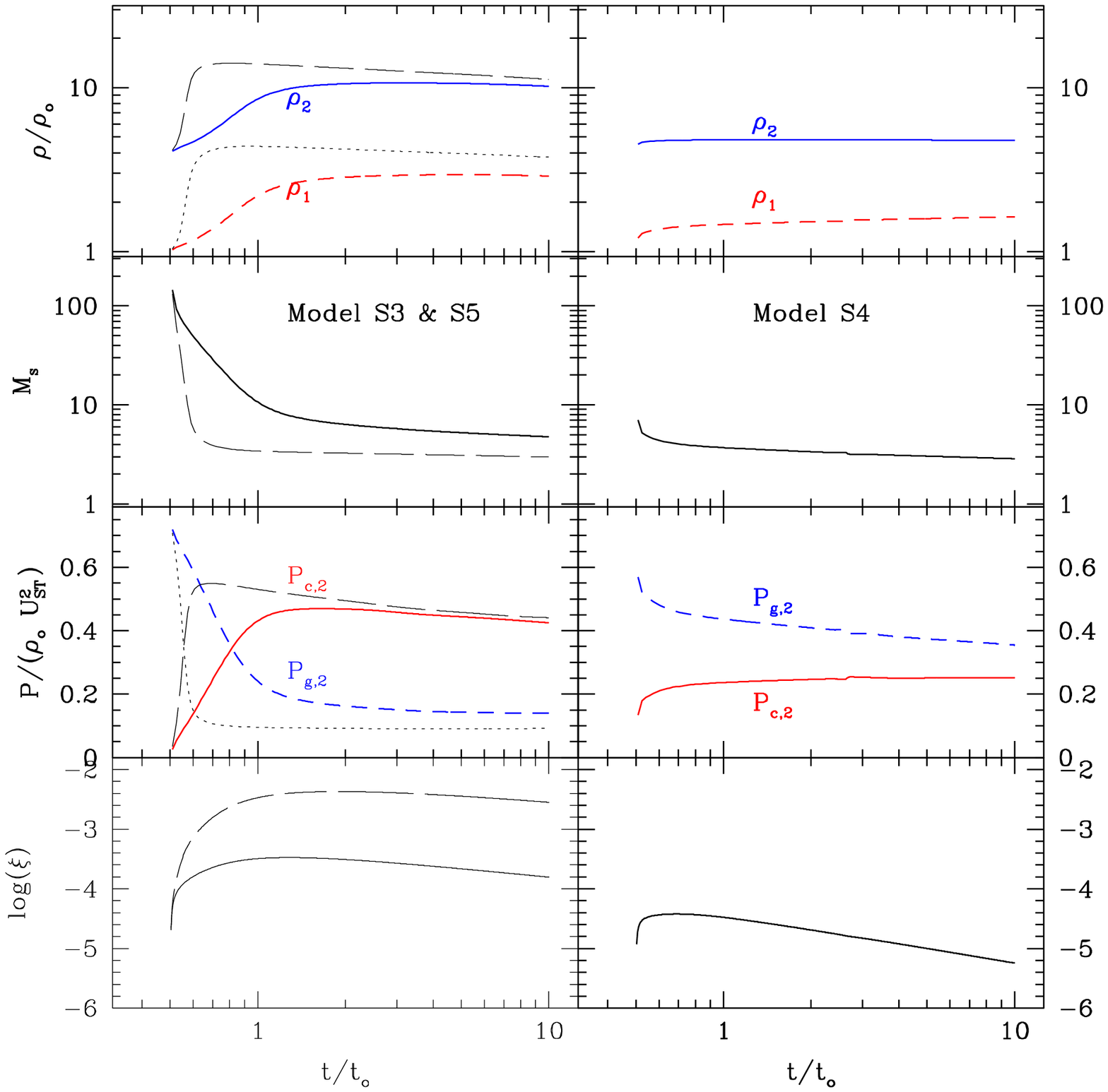}}
\vskip -0.5cm
\caption{
Pre-subshock density, $\rho_1$, post-subshock density,
$\rho_2$, shock Mach number, $M_s$, post-subshock CR and gas pressure
in units of the ram pressure of the unmodified Sedov-Taylor solution,
$\rho_0 U_{ST}^2 \propto (t/t_o)^{-6/5}$, and the CR injection parameter, 
$\xi$, are plotted for models S3 and S4.
The left panels also show model S5 with $\epsilon_B=0.2$
(long dashed lines and dotted lines).
}
\end{figure*}
\section{Results}
\subsection{SNR Model Parameters}
We consider a Type Ia supernova explosion with the ejecta mass, 
$M_{ej}=1.4 M_{\sun}$, in a warm or hot ISM with a uniform density.
Model parameters are summarized in Table 1.
The fiducial model, labeled S1 in Table 1, has the explosion energy, $E_o=10^{51}$ ergs, 
and the background density, $n_{\rm ISM}=0.3~ {\rm cm}^{-3}$.
The pressure of the background gas is set to be
$P_{\rm ISM} \approx 10^{-12}~{\rm erg ~cm^{-3}}$,
which determines the sound speed of the upstream gas and so
the Mach number of the SNR shock.
Recent X-ray observations of young SNRs indicate a magnetic field
strength much greater than the mean ISM field of $5\mu$G,
or values expected by compression of that field
(\eg Berezhko \etal 2003; V\"olk \etal 2005).
It is believed that the magnetic field upstream from the
shock is amplified 
by the CR streaming instability in the precursor region
(Bell 1978; Lucek \& Bell 2001).
Thus, to represent this effect we take $B=30\mu$G as the fiducial field strength.
The strength of magnetic field determines the size of
diffusion coefficient, $\kappa_n$, and the drift speed of Alfv\'en
waves relative to the bulk flow.
The Alfv\'en speed is given by $v_A= v_{A,0}(\rho/\rho_0)^{-1/2}$
where $v_{A,0} = (1.8 {~\rm km s^{-1}})B_{\mu}/\sqrt{n_{\rm ISM}}$. 
The second model, S2, assumes a higher explosion energy, while the
third model, S3, assumes the ISM magnetic field, rather than the 
amplified field. 
Model S4 assumes a hot ISM ($T\approx 10^6$ K), while all other 
models assume a warm ISM ($T\approx 10^4$ K).
For models S1-S4 $\epsilon_B=0.16$ is adopted. 
Model S5 has the same parameters as model S3 except
$\epsilon_B=0.2$, which allows a higher injection rate.

The physical quantities are normalized, both in the numerical code and in
the plots below, by the following constants:
\begin{eqnarray}
r_o=\left({3M_{ej} \over 4\pi\rho_o}\right)^{1/3},\nonumber\\
t_o=\left({\rho_o r_o^5 \over E_o}\right)^{1/2}, \nonumber\\
u_o=r_o/t_o,\nonumber\\ 
\rho_o = (2.34\times 10^{-24} {\rm g cm^{-3}}) n_{\rm ISM},\nonumber\\ 
P_o=\rho_o u_o^2.\nonumber
\end{eqnarray}
These values are also given in Table 1 for reference.

\begin{figure*}[t]
\vskip -0.5cm
\centerline{\epsfysize=14cm\epsfbox{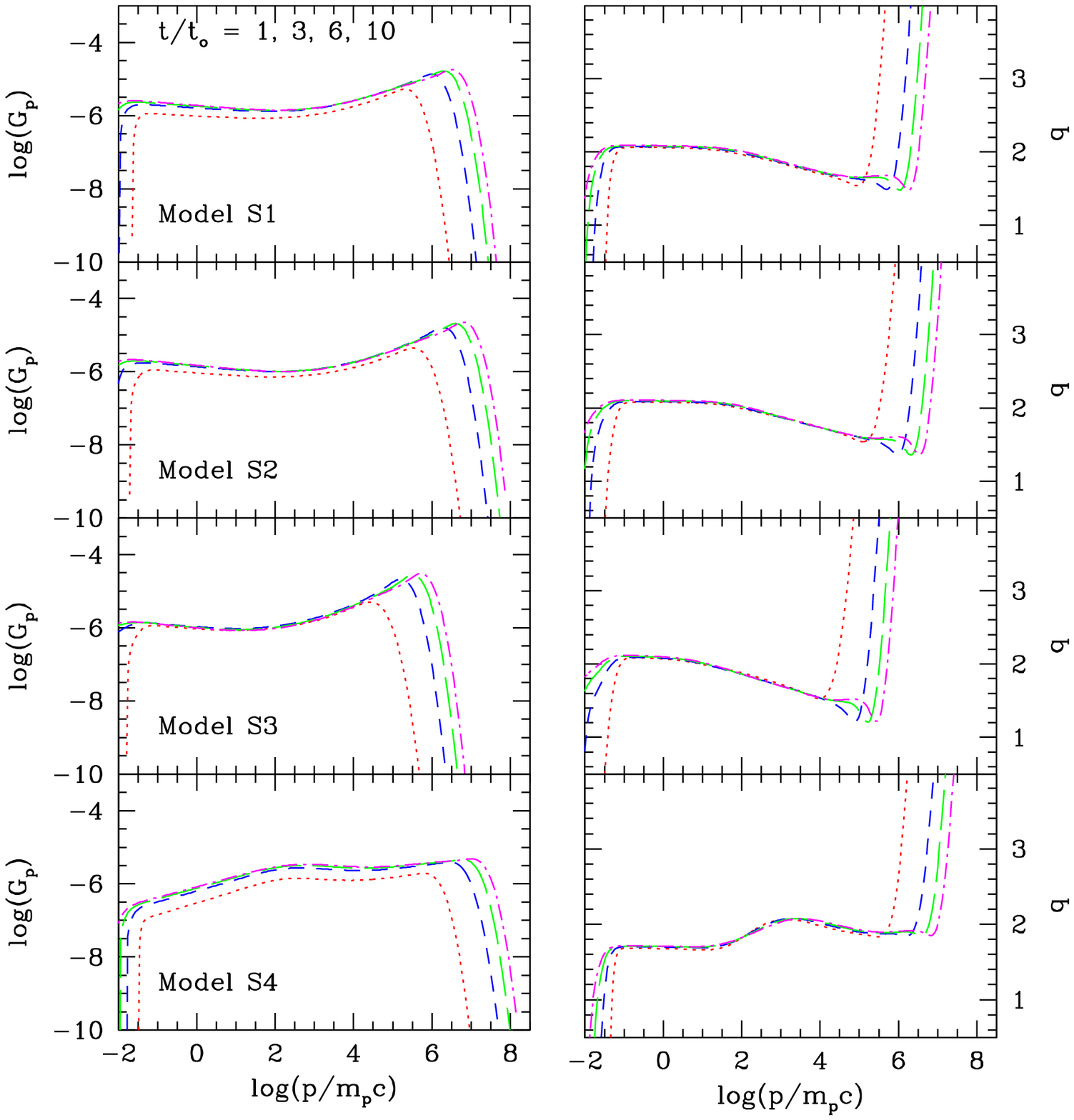}}
\vskip -0.5cm
\caption{
Integrated CR number, 
$G_p= \int g(p) r^2 {\rm d}r$, in arbitrary units and
the slope, $q = - d (\ln G_p)/ d \ln p - 2 $, are shown
at $t/t_o=$ 1, 3, 6,  and 10.
See Table 1 for model parameters for models S1-4.
}
\end{figure*}

\subsection{Remnant Evolution}

Fig. 2 shows the evolution of model S1 for $t/t_o = 0.5-10$.
The solid lines are for the initial condition 
which is assumed be the Sedov Taylor similarity solution
($R_{\rm ST}/r_o = \xi_s (t/t_o)^{0.6}$, 
$U_{\rm ST}/u_o= 0.6\xi_s (t/t_o)^{-0.4}$, where
$\xi_s=1.15167$) extrapolated to $t/t_o=0.5$.
Initially there is no pre-existing CRs and so all CR particles
are freshly injected at the shock.
The CR pressure becomes dominated over the gas pressure 
and the density compression across the total shock 
becomes $\rho_2/\rho_0 \approx 7$ after $t/t_o \sim 1$.
As the shock slows down and the CR pressure increases, 
the subshock Mach number decreases and 
the thermal population cools down, 
resulting in less efficient particle injection at low energies.
We note that the inner boundary of the simulation grid moves
out with the expanding comoving grid.

We repeated the same calculation starting from an earlier time,
$t/t_o=0.2$, to explore how the CR acceleration  
during the early free expansion stage would affect the results
at late ST stage.
The total CR energy accelerated by $t/t_o=10$ differs about 3 \%
and the CR spectrum extends to slightly higher $p_{\rm max}$ in
the simulation started from $t/t_o=0.2$, as one would anticipate
from the longer acceleration interval. The small differences
indicate that omitting the evolution before $t/t_o=0.5$
does not affect the overall conclusions of this work.

\begin{figure*}[t]
\vskip -0.5cm
\centerline{\epsfysize=14cm\epsfbox{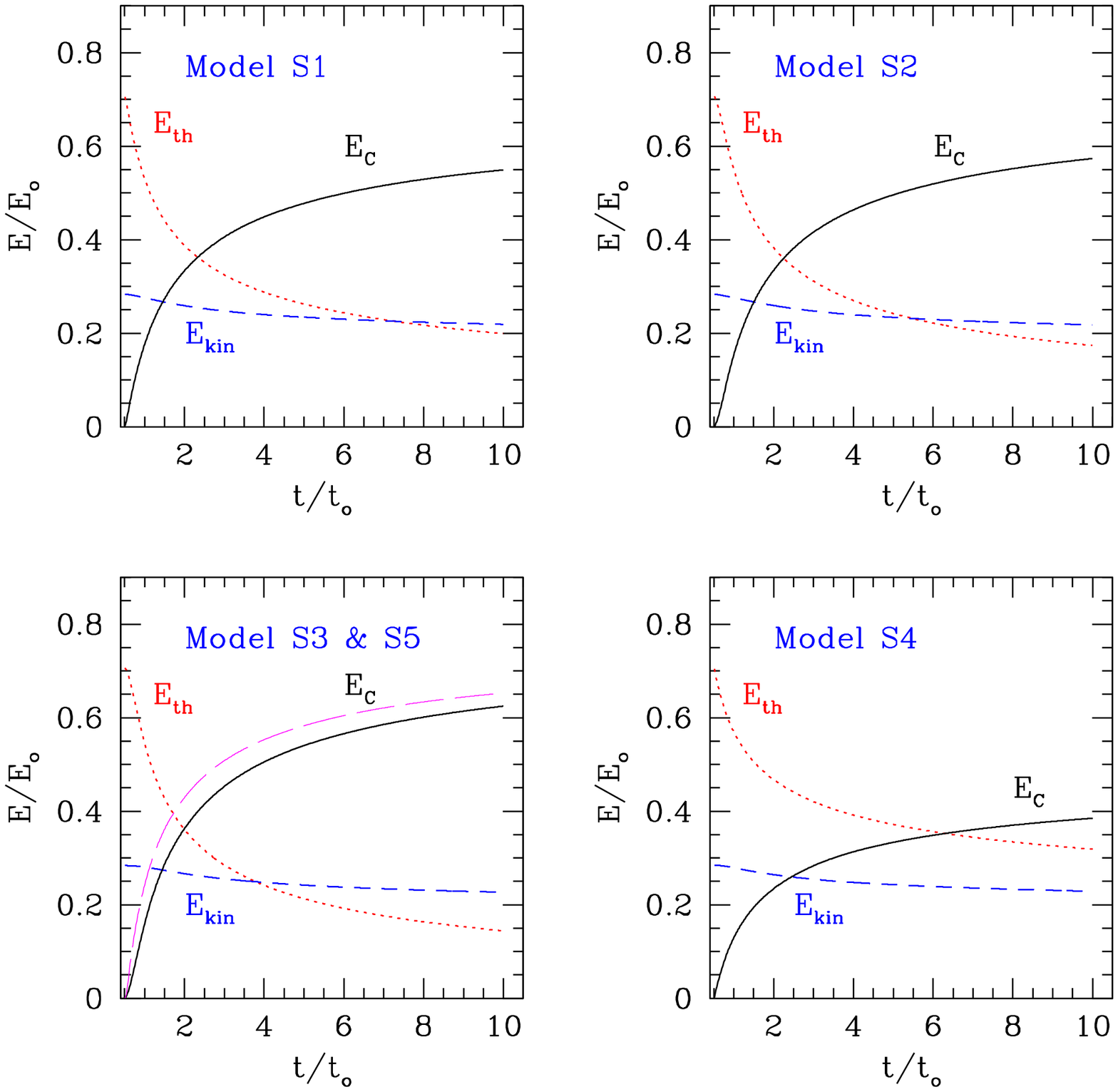}}
\vskip -0.5cm
\caption{
Integrated thermal, kinetic and CR energies inside the simulation volume
as a function of time.
See Table 1 for model parameters for models S1-S5.
The long dashed line in the lower left panel shows the volume
integrated CR energy for model S5.
}
\end{figure*}

\subsection{CR Injection and Acceleration}

The efficiency of the particle injection is quantified by
the fraction of particles swept through the shock 
that have been injected into the CR distribution:
\begin{equation}
\xi(t)=\frac {\int 4\pi r^2{\rm d r} \int 4\pi f_{\rm CR}(p,r,t)p^2 {\rm d p}}
{ \int 4\pi r_s^2 n_0 u_s {\rm dt }},
\end{equation}
where $f_{\rm CR}$ is the CR distribution function, $n_0$ is the
particle number density far upstream and $r_s$ is the shock radius. 

Figs. 3-4 show the evolution of shock properties such as
the compression factors, subshock Mach number, 
postshock pressures, and the injection fraction for all five models.
Most of these quantities approach to time-asymptotic values.   
The shock Mach number is the key parameter that determines the
CR injection and acceleration efficiency.
Model S4 has much lower initial Mach number,
so it shows less injection rate ($\xi \sim 10^{-5}$) compared
to the other models ($\xi \gsim 10^{-4}$).  
The compression factor in the precursor, $\rho_1/\rho_0\approx 2-3$,
while the compression factor across the total shock structure varies
somewhat, $\rho_2/\rho_0 \approx$ 7.0, 8.2, 10., and 4.8 
for models S1-S4, respectively. 
The postshock CR pressure relative to the ram pressure of the Sedov
solution is $P_{c,2}/\rho_0 U_{ST}^2 \approx$ 0.35, 0.39,
0.42, and 0.25 for models S1-S4, respectively.

We note that model S4 with $B=5\mu$G has the higher CR pressure than the
fiducial model S1 with $B=30\mu$G.
While the mode with higher B has smaller diffusion coefficient and 
so the CR spectrum extends to higher $p_{\rm max}$ (see Fig. 5),
the reduction of the injection and acceleration due to the wave drift 
leads to less CR energy.

Model S5 has a larger value of $\epsilon_B=0.2$ than model S3 does, 
but otherwise they have the same model parameters.
Weaker turbulence (larger $\epsilon_B$) lead to higher thermal leakage,
so the injection rate $\xi$ is about 15 times larger than that of
models S1-S3. This in turn results in slightly higher CR pressure 
and more significant precursor compression. However, it demonstrates that
the CR acceleration efficiency depends only weakly on the injection rate.

Fig. 5 shows the volume integrated CR spectrum,
$G(p) = \int 4\pi g(p) r^2 {\rm d}r$ in code units and 
the slope, $q = - d (\ln G_p)/ d \ln p - 2 $. 
The maximum momentum depends on the SNR parameters as
$p_{\rm max}\propto u_s^2 t/\kappa_n \propto B_\mu E_o \rho_{\rm ISM}^{-1/3}$.
The values of $\log(p_{\rm max}) \approx $ 6.5, 6.8, 5.7, and 7.0 at $t/t_o=10$
for models S1-S4, respectively, are roughly consistent with this relation. 

The CR spectra for models S1-S3 with high initial Mach numbers show the 
canonical nonlinear concave curvature for $p\gg 1$, which comes from
the large density compression factor ($\rho_2/\rho_0 \gg 4$) due to
the dominant CR pressure.
For these models the slope $q$ is close to 2 at low energies, 
but flattens to $\sim 1.6$ at high energies below the upper momentum cutoff.
For model S4 with lower initial Mach number the density compression is
only $\rho_2/\rho_0 \approx 4.8$, so the flattening of $G(p)$ at high
momenta is not so prominent.  
Instead, the decrease of the CR 
injection rate due to the weakening subshock reduces the particle
numbers at low momentum, leading to actually flatter spectra there.
 
Fig. 6 shows the integrated energies, 
$E_i/E_o = 4\pi \int e_i r^2 {\rm d}r$, where $e_{th}$,
$e_{kin}$, and $e_C$ are the density of thermal, kinetic and
cosmic ray energy, respectively.
The kinetic energy reduces only slightly and is similar for all models.
The total CR energy accelerated up to $t/t_o= 10$ is
$E_C/E_o=$ 0.55, 0.57, 0.62 and 0.39 for models S1-S4, respectively.
So models S2 (larger $E_o$) and S3 (smaller $B_{\mu}$) are a bit more
efficient in transferring the SN explosion energy to the CR energy,
compared to the fiducial model.
As mentioned earlier, comparison between models S1 and S3 indicates
that strongly amplified magnetic field carries faster Alfv\'en
waves and leads to less efficient CR acceleration, 
even though the particles are accelerated to higher energies.
Also comparison between models S3 and S5 shows that the total
CR energy increases only slightly, less than 3 \%, even though
the injection rate increases by a factor of about 15 with a larger
value of $\epsilon_B$ for model S5.
In model S4 with a hotter ISM, the shock is weaker and so the CR
injection and acceleration are less efficient. 

Our simulations imply that on average 
about $10^{-4}-10^{-3}$ of the incoming particles are injected
to the CR population at the shock front and  
up to 50 \% of the SN explosion energy can be
transferred to CRs during the ST stage,
if the magnetic field direction is radial (\ie quasi-parallel
field).
This last result is consistent with the calculations previously done 
by Berezhko and collaborators using a different numerical scheme
(\eg Berezhko \& V\"olk 1997).
The CR injection rate, however, probably depends strongly on the angle
between the magnetic field and the shock normal direction.
In a more realistic magnetic field geometry, where a uniform
ISM field is swept by the spherical shock, only 10-20 \% of the
shock surface has a quasi-parallel field geometry (V\"olk \etal 2003).
In the shock surface region where the field is perpendicular,
the injection rate is expected to be reduced and so the CR acceleration
efficiency would be smaller. 
Thus the CR energy conversion factor averaged over the entire
shock surface could be significantly smaller than 50 \%,
perhaps about 10 \%.
On the other hand, Giacolne (2005) showed that the protons can
be injected efficiently even at perpendicular shocks in fully
turbulent fields due to field line meandering.
In such case the injection rate at perpendicular shocks may 
not be significantly smaller, compared to parallel shocks. 

\section{SUMMARY}

The evolution of cosmic ray modified shocks depends on complex
interactions between the particles, waves in the magnetic field, 
and underlying plasma flow.
We have developed numerical tools that can emulate some of those
interactions and incorporated them into a standard numerical 
scheme for gasdynamic problems.
Specifically, diffusive shock acceleration can be followed by
a kinetic approach (\ie CRASH code) 
in which a diffusion convection equation for the
CR distribution function is solved  with an appropriate diffusion 
coefficient (Kang \& Jones 1991; Kang \etal 2001).
The injection of CR particles can be treated by a thermal leakage
injection model with a transparency function
$\tau_{\rm esc}(\epsilon_B, \upsilon)$ (Kang \etal 2002).
Drifts of resonantly-scattering Alfv\'en waves and heating of the
thermal plasma due to the dissipation of those waves in the precursor 
region has been included through a simple model 
(Jones 1993; Kang \& Jones 2006).

In the present paper, we applied the spherical CRASH code
to the problem of the CR acceleration at
the remnant shock from typical Type Ia SNe
propagating into a uniform interstellar medium. 
The main results of our simulations can be summarized as follows:

1) Since the CR injection and acceleration depend primarily on the
shock Mach number, the temperature of the ISM is an important
factor. SNRs in a warm ISM inject about $10^{-4}-10^{-3}$ of the particles
passed through the shock and transfer about
50\% of the explosion energy to the CR component. 
SNRs in a hot ISM inject 10 times less particles, but
they still transfer about 40 \% of the explosion energy
to CRs. 

2) At sources, the hadronic CR spectrum should be consistent with
a power-law function, $N(E)dE \propto E^{-2}dE$ up to $p/m_{\rm p}c\sim 10^2$, 
but it may gradually flatten to $E^{-1.6}$ toward the Knee energy and
beyond up to $E\sim 10^{16} Z$ eV ($p/m_{\rm p}c\sim 10^7$).  

3) About 40-50 \% of the explosion energy can be transfered to
CRs, if the magnetic field is radial (\ie quasi-parallel shocks). 
However, more realistic
consideration of the field geometry relative to the spherical 
shock surface could lead to much smaller energy conversion rate.
So a conservative estimate could be order of 10 \%.  

4) If the magnetic field is amplified in the precursor region
due to the streaming instability, as indicated by recent X-ray
observations of young SNRs, 
the particles are accelerated to higher energies due to smaller
diffusion coefficient. However, faster Alfv\'en waves 
in the precursor tend to
reduce the CR injection and acceleration efficiencies. 

In conclusion, the galactic cosmic rays possibly up to $10^{16}Z$ eV  
could be originated from supernova remnants. 
This result is quite robust, regardless of SNR model parameters
and details of microphysics involved in the injection process,
provided that the Bohm diffusion is valid at young SNRs
of several thousand years old.

\acknowledgments{
This work was supported for two years by Pusan National University 
Research Grant. The author would like to thank T.W. Jones for helpful
comments on the paper.}

\end{document}